\documentclass[12pt,preprint]{aastex}

\accepted{}

\slugcomment{to appear in {\it The Astrophysical Journal}}

\lefthead{Kraemer et al.}
\righthead{T} 

\received{2003 August 6}
\begin{document}

\title{On the Relationship Between the Optical Emission-Line and X-ray Luminosities in Seyfert 1
Galaxies}

\author{S. B. Kraemer\altaffilmark{1},
I.M. George\altaffilmark{2,3},
D. M. Crenshaw\altaffilmark{4},
\& J.R. Gabel\altaffilmark{1}}

\altaffiltext{1}{Catholic University of America,
and Laboratory for Astronomy and Solar Physics, NASA's Goddard Space Flight 
Center, Code 681,
Greenbelt, MD  20771; kraemer@yancey.gsfc.nasa.gov, gabel@iacs.gsfc.nasa.gov}

\altaffiltext{2}{Joint Center for Astrophysics, Physics Department, University of Maryland,
Baltimore County, 1000 Hilltop Circle, Baltimore, MD 21250.}

\altaffiltext{3}{Laboratory for High Energy Astrophysics, NASA's Goddard Space Flight
Center, Code 662, Greenbelt, MD  20771:
ian.george@gsfc.gsfc.nasa.gov}

\altaffiltext{4}{Department of Physics and Astronomy, Georgia State University,
Atlanta, GA 30303; crenshaw@chara.gsu.edu}

\begin{abstract}

We have explored the relationship between the [O~III] $\lambda$5007 and the 2--10 keV luminosities for a sample of Broad- and Narrow-Line
Seyfert 1 galaxies (BLSy1 and NLSy1, respectively). We find that both types of Seyferts 
span the same range in luminosity and possess similar [O~III]/X-ray ratios. The NLSy1s are more luminous than 
BLSy1s, when normalized to their central
black hole masses, which is attributed to higher mass accretion rates. However, we find no evidence for elevated [O~III]/X-ray ratios
in NLSy1s, which
would have been expected if they had excess  
EUV continuum emission compared to BLSy1s. Also, other studies suggest that the gas in narrow-line regions (NLR)
of NLSy1s and NLSy1s span a similar range in ionization, contrary to what is expected if those of the former
are exposed to a stronger flux of EUV radiation. The simplest interpretation is that, like BLSy1s, a large EUV bump is not 
present in NLSy1s. However, we show that the [OIII]/X-ray ratio can be lowered as a result of absorption of the ionizing continuum by gas 
close to the central source, although there is no evidence that intrinsic line-of-sight absorption is more common 
among NLSy1s, as would be expected if there were a larger amount of circumnuclear gas. Other  
possible explanations include: 1) anisotropic emission of the ionizing radiation, 2) higher gas densities in the
NLR of NLSy1s, resulting in lower average ionization, or 3) the  
presence of strong winds in the the nuclei of NLSy1s which may drive off much of the gas in the narrow-line
region, resulting in lower cover fraction and weaker [O III] emission. 

\end{abstract}


\section{Introduction}

Osterbrock \& Pogge (1985) discovered a class of Seyfert 1 galaxies characterized by relatively narrow
($FWHM$ $\leq$ 2000 km s$^{-1}$) permitted lines, which they dubbed ``Narrow-Line Seyfert 1s''(NLSy1s).
Compared to their broad-line counterparts (BLSy1s), the X-ray continua of NLSy1s are characterized by steeper slopes in the soft
and hard bands (Boller, Brandt, \& Fink 1996; Brandt, Mathur, \& Elvis 1997) and more rapid variability (Turner et al. 1999).
In a {\it ROSAT} sample of active galaxies (AGN), Grupe et al. (1998) found similar optical to X-rays spectral indices for BL- and NLSy1s, which suggests 
that the soft X-ray steepness is 
indicative of excess emission, rather than weakness in the hard X-ray band.
They also found that the optical and soft-X-ray spectral indices were anti-correlated
suggesting that their continua peak in the EUV. There have been several models proposed to explain the more extreme properties of NLSy1s. 
For example, Wandel, Peterson, \& Malkan (1999)
suggested that NLSy1s possess a more distant Broad Line Region, due to over-ionization. Another
possibility is that the BLR is flattened and
viewed roughly face-on, with narrow line profiles as a result (e.g. Boller at al. 1996). The currently popular paradigm
posits that, while AGN in general are powered by
accretion of material onto a supermassive central black hole, NLSy1s possess black holes
of relatively modest mass ($\leq$ 10$^{7}$ M$_\odot$), accreting matter at or above
its Eddington limit (Pounds, Done, \& Osborne 1995). The narrow line widths are then due to clouds
in motion around the small-mass black hole, while the steep soft X-ray continuum is the 
high energy tail of the ``Big Blue Bump'', which is presumably emission from the accretion
disk. Due to the high accretion rates (\.{M}) , the disks in NLSy1s are likely much hotter than those
in BLSy1s, hence the emission is peaked at higher energies. Wang \& Netzer (2003) showed
that high \.{M} results in an extremely thin disk which emits a
double-peaked continuum, with a peak in the EUV-soft X-ray due to the disk itself, and
a high energy peak due to a hot corona, and, further, that most 
NLSy1s are super-Eddington accretors. 

Wills et al. (1999) found unusually high N~V $\lambda$1240/ C~IV $\lambda$1550 flux ratios in
NLSy1s, which may be indicative of high nitrogen abundances, as would result from metal
enrichment due to a recent burst of star formation (see, also, Shemmer \& Netzer 2002). This may indicate a different 
circumnuclear environment in NLSy1s compared to BLSy1s. Kuraszkiewicz et al. (2002)
compared the UV spectra of NLSy1s and BLSy1s and determined that the former are characterized by weaker C~IV $\lambda$ 1549 and
C~III] $\lambda$1909 and stronger Al~III $\lambda$1857 emission. They suggested that the emission-line gas
is of higher density and lower ionization than BLR gas in BLSy1s (see, also,
Marziani et al. 2001). 
It has been noted (Mathur 2000, and references therein) that NLSy1s possess strong Fe~II and weak [O~III] $\lambda$5007 emission relative to 
H$\beta$. This puts them at one extreme of eigenvector 1 of Boroson \& Green (1992), derived
from principal component analysis of a large sample of low-redshift QSOs.
In this regard, the NLSy1s do tend to show stronger optical Fe~II emission than BLSy1s (e.g., Gaskell 1985; Zheng \& Keel 1991), however
it has recently been suggested that this is a result of weaker H$\beta$ emission (Veron-Cetty, Veron,\& Goncalves 2001). On the other hand,
by deconvolving the
broad and narrow components of the H$\beta$ profiles, Veron-Cetty et al. (2001)
argued that the the ratio of [O~III] to narrow H$\beta$ was similar to that of BLSy1s, which suggests that conditions
in the Narrow-Line Regions (NLR), i.e, where the forbidden emission lines arise, are similar among the two groups of Seyferts. 
If the ionizing continuum in NLSy1s peaks more strongly in the EUV compared to BLSy1s, one would expect stronger emission
from ions with high ionization potentials, such as [Fe~VII] $\lambda$6087
and [Fe~X] $\lambda$6374, relative to [O~III] $\lambda$5007, however there is no strong
evidence for this effect (Nagao et al. 2000). Furthermore, although Rodriguez-Ardila et al. (2000) found that
the NLR of NLSy1s may be more highly ionized (or lacking some low-ionization gas) and that the [O~III] 
emission-line gas is somewhat denser than in BLSy1's, the two classes span the same range in [O~III]
luminosity. Note, however, that these studies did not fully explore the relationship between the emission lines
and the luminosity of the AGN. In this paper, we compare the luminosities of the [O~III] $\lambda$5007 (L$_{[OIII]}$) to that 
in the 2--10keV band (L$_{2-10}$) for a sample of BLSy1s and NLSy1s, in order to determine if there is evidence for
a stronger EUV flux in the latter. Specifically, if the conditions (density,
cover factor, etc.) in the NLRs of BL- and NLSy1s are not significantly different, the [O~III] emission will be
a reliable probe of their luminosities in the EUV, with higher EUV luminosities producing more [O~III] per
hard X-ray luminosity.     

\section{Luminosity Comparison}

In selecting the sample of BLSy1s and NLSy1s, we took those objects with available black hole mass
estimates and/or 2-10 keV fluxes and assembled their [O~III] fluxes from the literature. 
The values for each object in the sample are listed in Tables 1 and 2, along with the references for the line fluxes and black hole masses.
We did not correct the emission-line fluxes for reddening. The X-ray fluxes were from {\it ASCA} observations (with one exception) and were
retrieved from the Tartarus database (Turner et al. 2001). The fluxes were not corrected for Galactic 
or intrinsic absorption. In the case of multiple observations, we used the mean flux.
Fluxes were converted to luminosity assuming H$_{0}$ $=$ 75 km s$^{-1}$ Mpc$^{-1}$ and using 
redshifts from NASA/IPAC ExtraGalactic Database ($http://nedwww.ipac.caltech.edu/$). Clearly, this is not a complete sample in any sense, however with 26 BLSy1s and 18 NLSy1s
there
are a sufficient number of data points that we were able to assess the statistical robustness of relationships 
among physical quantities (within
the sample). 
   
In Figure 1, we show L$_{[OIII]}$ versus L$_{2-10}$ for both types of Seyferts.
The fact that these data were not corrected for extinction or X-ray absorption is the cause of
some of the scatter in the plot. To demonstrate the magnitude of the effect, we marked the position
of three Seyferts in Figure 1. MCG-6-30-15 is known to be highly reddened (e.g. Brandt, Fabian, \& Pounds 1996), which results
in a lower L$_{[OIII]}$ compared to L$_{2-10}$, as is clearly seen by its position below that of most 
other galaxies. NGC 4151 has an X-ray absorber with a large column density ($>$ 10$^{22}$ cm$^{-2}$; e.g. George et al. 1998), but little intrinsic
reddening (Kriss et al. 1992), hence it lies somewhat to the left (lower X-ray luminosity) than those objects
with similar L$_{[OIII]}$. Finally, NGC 3227 is both reddened and possesses a large absorbing column (Kraemer et al.
2000a, and references therein), and the combination of effects drive it to the lower left in the diagram. However,
the scatter is not sufficient to obscure the relationship between these quantities. Another possible source of scatter is
variability in the X-ray band, and there has not been extensive monitoring for the majority of these
sources. However, it is unlikely that the mean fluxes are off by more than a factor
of a few from the true intrinsic fluxes, which is not sufficient to destroy any relationship.  

On first inspection, it appears that the two classes 
occupy the same region in the diagram. To determine if they differed, we first applied a Kolomgorov-Smirnov (K-S) statistical test to the values of
L$_{2-10}$ for the two types of Seyferts. While the NLSy1 sample is weighted to somewhat higher redshift, which might
lead one to think that it would be skewed to higher luminosity, the test returned a 73\% probability of the
null hypothesis (i.e., that they were from the same population). We next applied
a K-S test to the ratios of L$_{[OIII]}$ to L$_{2-10}$. The test
returned an 89.4\% probability of the null hypothesis. For comparison,
we applied a K-S test to the BLSy1s and a set of the BLSy1 values randomly adjusted up or down by 5\%, which
returned an 86\% probability. Hence, both BLSy1s and NLSy1s exhibit the same relation between L$_{[OIII]}$ and L$_{2-10}$.

To explore the physical connection between these quantities, we generated single-zoned models for the [O~III] emission-line gas
using the photoionization code Cloudy (Ferland et al. 1998). We assumed typical conditions for the [O~III] emitting region
in the NLR, e.g. a hydrogen number density of 10$^{4}$ cm$^{-3}$ (see Kraemer et al. 2000b; Kraemer \& Crenshaw 2000), a column density of 10$^{21}$ cm$^{-2}$ 
(which ensured that the models were radiation bounded), and  
solar abundances (e.g. Grevesse \& Anders 1989).
For the sake of simplicity,
we assumed that the gas was free of cosmic dust. We adopted two different forms for the spectral
energy distribution (SED) of the ionizing continuum. First, we assumed broken power-law similar to that
suggested for NGC 5548 (Kraemer at al. 1998) and NGC 4151 (Alexander et al. 1999; Kraemer et al. 2000b), of the
form F$_{\nu}$ $\propto$ $\nu$$^{-\alpha}$, with $\alpha$ $=$ 0.5 below 13.6 eV, 1.5 from 13.6 eV to 1.0 keV, and 0.8 at higher energies.
For the second SED, we included EUV excess by adding a steeper soft-X-ray component extrapolated to meet the UV continuum at 100 eV, with the  
following spectral indices: $\alpha$ $=$ 0.51 below 100 eV, 2.0 from 100 eV to 1.0 keV, and 0.8 at higher energies.
The two SEDs are shown in Figure 2. To ensure that the second SED represents the general characteristics of the
soft X-ray continuum of NLSy1s, we have overlain the 0.25 keV, 1 keV, and 2 keV luminosities
for the NLSy1s Ton S180 (Turner et al. 2002) and Ark 564 (Romano et al. 2004; Matsumoto, Leighly, \& Marshall 2004), normalized to their reddening-corrected
1000 \AA~luminosities (note that the 0.25 keV point for Ark 564 was determined by extrapolation, since the data only 
extend to 0.5 keV). Both of these sources show the steep soft X-ray continua typical of NLSy1s (see Boller et al. 1996), and
our model continuum shows the same overall slope, lying between the two sets of datapoints. Obviously, 
we are not trying to model the physical nature of the soft X-ray emission, which may result from a combination of 
black-body and power-law components (see Turner et al. 2002), but rather using it to illustrate the effects of an EUV excess.
As such, although NLSy1s are known to possess steeper hard X-ray continuua than BLSy1s (Brandt et al. 1996), we have used the same 
2-10 keV indices for both types, since this allows us to directly examine the effect of the EUV excess. 
   
For the BLSy1 model, we assumed an ionization parameter (number of ionizing photons per nucleon at the illuminated face of the zone)
of $U = 0.001$, which yields a predicted [O~III]/H$\beta$ ratio of $\approx$ 10, typical for the NLR of Seyfert 1s (Cohen 1983). 
For the BLSy1 SED model and the average X-ray luminosity of the sample,
1.4 x 10$^{43}$ erg s$^{-1}$, the luminosity in ionizing photons is 1.7 x 10$^{54}$ photons s$^{-1}$,
which places the gas at $\sim$ 200 pc from the central source. Fixing the 13.6 eV and 2 keV fluxes to their BLSy1 model values, the NLSy1
SED increased $U$ by a factor of 2.15. The model (the ``high U'' NLSy1 model) predicts [O~III]/H$\beta$ and L$_{[OIII]}$/L$_{2-10}$ ratios greater by factors of 2 and 5, respectively,
than the BLSy1 model. Note, however, that there is no evidence that the [O~III]/H$\beta$ ratios are higher in NLSy1s than in
BLSy1s (Rodriguez-Ardila et al. 2000; Veron-Cetty et al. 2001). Therefore, we ran a third model, using the NLSy1 SED but setting $U = 0.001$.
This model (the ``low U'' NLSy1 model) predicts smaller increases in the [O~III]/H$\beta$ and L$_{[OIII]}$/L$_{2-10}$ ratios, i.e., factors of 1.5 and 1.8, respectively.

In order to illustrate how the L$_{[OIII]}$ might scale with L$_{2-10}$, we used the ratios from each model and assumed
a covering factor of 0.02, as Netzer \& Laor (1993), with the resulting overlays shown in Figure 1. Note that implicit in the scaling of the model predictions is
the assumption that $U$ does not vary with the luminosity of the ionizing continuum, hence the more luminous objects must have more extended NLRs
and/or a contribution from denser gas. The scaled ratios L$_{[OIII]}$/L$_{2-10}$ are 0.010, 0.018, and 0.050, for the BLSy1, ``low U''
NLSy1,and ``high U'' NLSy1 models, respectively. The BLSy1 model fits the data reasonably well and follows the observed points over three orders
of magnitude in hard X-ray flux, hence there is no evidence for a narrow-line ``Baldwin Effect'' (e.g., Baldwin, Wampler, \& Gaskell 1989) in this luminosity range, unlike that observed
among higher luminosity AGN (Croom et al. 2002). The model prediction for the NLSy1 SED with the scaled ionization parameter lies well-above the majority of the 
data points, while the prediction for the model assuming $U = 0.001$ lies much closer to the BLSy1 result. Hence, if the ionization parameters
are indeed similar, {\it contrary to what might be expected with the stronger EUV flux in NLSy1s}, it is possible that the two classes
would be indistinguishable due to inherent scatter in the L$_{[OIII]}$/L$_{2-10}$ ratios resulting from intrinsic
reddening and X-ray absorption.  The implication is that
the physical conditions in the NLRs of NLSy1s and BLSy1s are, on the average different, e.g. NLSy1s have stronger emission from dense gas, in which 
the [O~III] line is
collisionally suppressed, or a more distant NLR. Rodriguez-Ardila et al. (2000) favor the former possibility.
in which case the overall NLR conditions. 

Another mechanism for decreasing the characteristic NLR ionization parameters in NLSy1s is absorption of the ionizing
radiation by gas close to the active nucleus (e.g. Kraemer et al. 1999, Alexander et al. 1999; Kraemer et al. 2000b). To illustrate this, we generated a filtered
ionizing continuum by modifying our NLSy1 SED via transmission through an absorber of ionization parameter $U = 0.005$, and total column density
of 2.5 x 10$^{20}$ cm$^{-2}$. These parameters were chosen to drop the ionization parameter, for the conditions
assumed in the higher ionization NLSy1 model, from 0.00215 to 0.001, and are similar to the strong components of intrinsic
UV absorption detected in NGC 4151 (Kraemer et al. 2001) and NGC 3516 (Kraemer et al. 2002). The filtered continnum is shown
as the dotted line in Figure 2, and is clearly optically thick just above the H and He~II Lyman limits. The resulting NLR
model predicts an [O~III]/H$\beta$ ratio within $\sim$ 10\% of the ``low U'' NLSy1 model, and L$_{[OIII]}$/L$_{2-10}$
$=$ 0.019, assuming the same NLR covering factor to scale the other model predictions.
Hence, the similarity in the [O~III]/X-ray ratios of BLSy1s and NLSy1s could result from the presence of larger amounts of circumnuclear gas in the 
latter. Note that the density of the intervening absorber must be high enough, i.e. $\gtrsim$ 10$^{6}$ cm$^{-3}$ ((Osterbrock 1989),
that the absorber does not produce significant [O~III] emission. Although the presence of such an absorber could be a consequence of the 
higher mass accretion rates suggested for
NLSy1s, there is no direct evidence for excess circumnuclear gas in NLSy1s. Specifically, NLSy1 do not posses stronger 
or a higher incidence of line-of-sight absorption than BLSy1s (see Crenshaw et al. 1999).

It is widely believed that, for a given luminosity, NLSy1s possess lower mass central black holes than BLSy1s. The black hole masses (M$_{bh}$) listed in Tables 1 and 2 were
determined by a variety of methods, including reverberation mapping (Wandel et al. 1999; Shemmer et al. 2001), 
the BLR radius/Luminosity relation (Kaspi et al. 2000; Wandel 2002, Bian \& Zhao 2003), and via the X-ray power density
spectra (Czerny et al. 2001). It must be noted that the masses derived from the BLR radius/Luminosity relation depend on line width,
and there may be systematic differences between the values determined for NLSy1s compared to BLSy1s. Hence,
one must be circumspect in drawing conclusions from such a heterogeneous set of estimated masses.
In Figure 3, we show L$_{[OIII]}$ as a function of M$_{bh}$, which 
shows the same properties as that of X-ray luminosity versus M$_{bh}$ (see Figure 1 in Alonso-Herrero et al. [2002]).
It is readily apparent that NLSy1s and BLSy1s occupy different regions; a K-S
test of the quantity L$_{[OIII]}$/M$_{bh}$ returned a 1.6\% probability of the null hypothesis. Furthermore, Ark 564, Mrk 110, Mrk 335, and NGC 4051,
for which the values of M$_{bh}$
were determined via reverberation mapping and are generally considered more reliable, occupy the roughly same region of Figure 2 as the
other NLSy1s, which implies that the observed relationship of L$_{[OIII]}$ and M$_{bh}$ is not obviously biased by
problems with the BLR radius/Luminosity relation. 

Assuming similar physical conditions in the emission-line gas, more [O~III] emission indicates a stronger flux of ionizing radiation.
Clearly, if the central engines in both BLSy1s and NLSy1s are powered by accretion of matter with similar efficiencies then, for a given M$_{bh}$, higher \.{M} 
will result in more
radiation. Furthermore, it is possible that high accretion rates also produce hotter disks (Wang \& Netzer 2003), hence more ionizing
radiation, and the differences in L$_{[OIII]}$ as a function of M$_{bh}$ for BLSy1s and NLSy1s shown in Figure 3 can be simply
interpreted in light of such models. Second, there is some overlap between the BLSy1s and NLSy1s, which may indicate
that there is a range of \.{M}/M$_{bh}$ among Seyferts. For the sake of comparison, we have overplotted L$_{[OIII]}$ as a function of M$_{bh}$, assuming 
L$_{bolometric}$/L$_{2-10}$ $=$ 27.2 (Awaki et al. 2001), and the scaling of L$_{[OIII]}$/L$_{2-10}$ described above, for Eddington ratios 
(L/L$_{edd}$) of unity and 0.01. As has been often noted (e.g. Wang \& Netzer 2003, and references therein), a number of NLSy1s appear to be radiating at 
close to L$_{edd}$.

\section{Discussion}

As illustrated by the K-S tests, our sample of BLSy1 and NLSy1s exhibit similar distributions in X-ray luminosity (which, of course,
is partly a selection effect, since we did not include LINERS and low-luminosity AGN in the sample), and we find a 
similar relationship between the [O~III] and hard X-ray luminosities.
If the ionizing SEDs of NLSy1s were
significantly different than BLSy1s, in the sense that they peaked in the EUV (e.g., Kuraszkiewicz et al. 2000), one would expect that
they would have larger luminosities in ionizing photons compared to their hard X-ray luminosities. As a result, one would also 
expect higher emission-line luminosities relative to the hard X-ray, which is not the case. Hence, the simplest interpretation of these
results is that, like BLsy1s (Laor et al. 1997),
NLSy1s lack a strong EUV bump, contrary to the observational evidence (Boller et al. 1996). However, the effect of an EUV excess could be 
masked by other differences between BLSy1s and NLSy1s.
For example, we have demonstrated that the presence of an intervening absorber which is optically thick above the He~II 
Lyman limit could modfiy the ionizing continuum sufficiently
to drop the average ionization in the NLR of NLSy1s. This could explain why the NLR in NLSy1s does not appear to more ionized
than that of BLSy1s (e.g. Nagao et al. 2000), and could be a consequence of high mass accretion rates. However, there is no 
direct evidence for a greater amount or distribution of circumnuclear
gas in NLSy1s required by this scenario. Furthermore, although some unabsorbed NLSy1s lie close to the [O~III]/X-ray line from the
high ionization NLSy1 model in Figure 1 (e.g. Mrk 478; Crenshaw et al. [1999]),
Ton S180, which shows little evidence for intrinsic absorption (Turner et al. 2001), lies well below the BLSy1 model line. 

A denser or more distant NLR in NLSy1s could also effectively mask the effects of
a stronger EUV flux, however, there is no strong evidence for either of these conditions.  
A number of NLSy1s show blueshifted [O~III] relative to H$\beta$ (e.g., I Zw 1; Phillips 1976) which may indicate strong winds in the NLR. This may 
result in lower NLR covering factors, which could drive the L$_{[OIII]}$/L$_{2-10}$ ratio down, but it is not clear how this might
affect the average ionization of the gas, since the lower ionization cloud only occur if the emission-line clouds were confined such that
their densities did not drop too rapidly with increasing radial distance. Another possible
explanation is that the NLR gas in the NLSy1s is much more highly ionized that than in BLSy1s, with a smaller fraction of doubly-ionized oxygen. 
However, if this were the case, one would expect stronger
high ionization lines, such as [Fe~X] $\lambda$6374, and Nagao et al. (2000) did not find strong evidence of this. 
A number of NLSy1s show blueshifted [O~III] relative to H$\beta$ (e.g., I Zw 1; Phillips 1976) which may indicate strong winds in the NLR. This may 
result in lower NLR covering factors, which could drive the L$_{[OIII]}$/L$_{2-10}$ ratio down. 
Finally, there is the possibility that the ionizing radiation is emitted anisotropically, hence the 
NLR gas is not exposed to the excess EUV radiation. However, there is no indication that the characteristics of NLSy1s can be the
result of preferential viewing angle (Boller et al. 1996)

Based on this sample, the strength of L$_{[OIII]}$ relative to M$_{bh}$ supports the idea that there is a range of Eddington
ratios among Seyfert 1s, with some of the NLSy1s occupying one extreme. 
While these results strongly suggest that the [O~III] emission is not unusual in NLSy1s (see, also, Veron-Cetty et al. 2001),   
perhaps other lines can be used to constrain the EUV continuum and, ultimately, the structure of the central engine. In fact, Veron-Cetty et al. 
(2001) suggest that it might be more advantageous to classify Seyferts based on the Fe~II emission. Clearly, more data are needed to
probe the nature of the NLSy1 phenomenon. For example, a larger set 
optical and UV spectra will be invaluable for probing any
differences in the emission-line gas that may not be apparent from [O~III] alone, and high resolution UV 
and X-ray spectra are required to help determine if the NLSy1 possess stronger intrinsic absorption.

\acknowledgments

 S.B.K. and D.M.C. acknowledge support from NASA grant NAG5-13109. 
 This research has made use of the Tartarus database, which is supported by 
 Jane Turner and Kirpal Nandra under NASA grants NAG5-7385 and NAG5-7067.

\clearpage
\figcaption[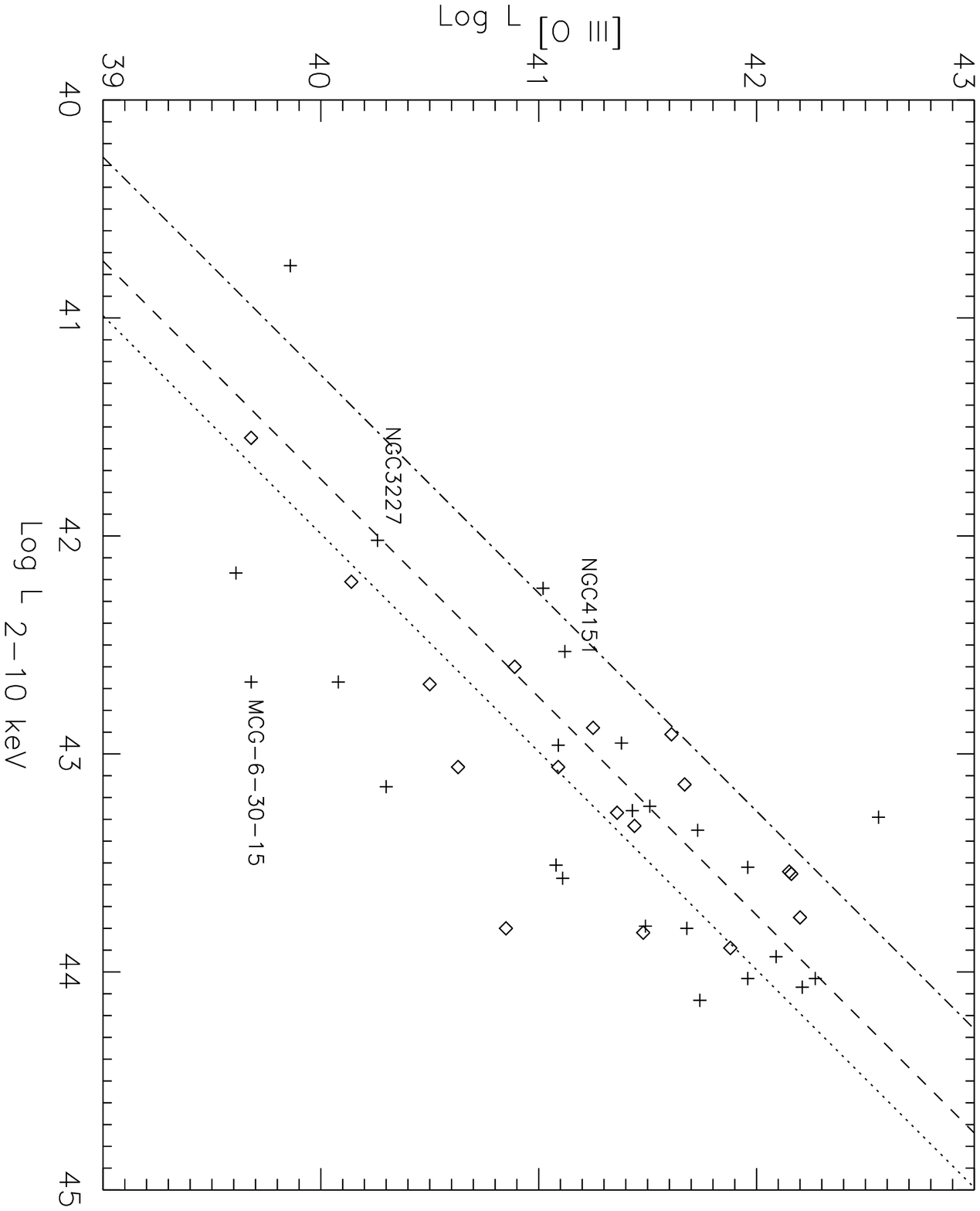]{A comparison of the [O~III] $\lambda$5007 and
2--10 keV band luminosities for the Broad-Line (crosses) and Narrow-Line
(diamonds) Seyfert galaxies in the sample. The dotted (BLSy1 SED, $U = 0.001$), 
dash-dotted (NLSy1 SED, $U = 0.00215$), and dashed (NLSy1 SED, $U = 0.001$) lines show the
photoionization model predictions for [O~III] as a function of the 2-10 keV luminosity 
(see the text for details). }

\figcaption[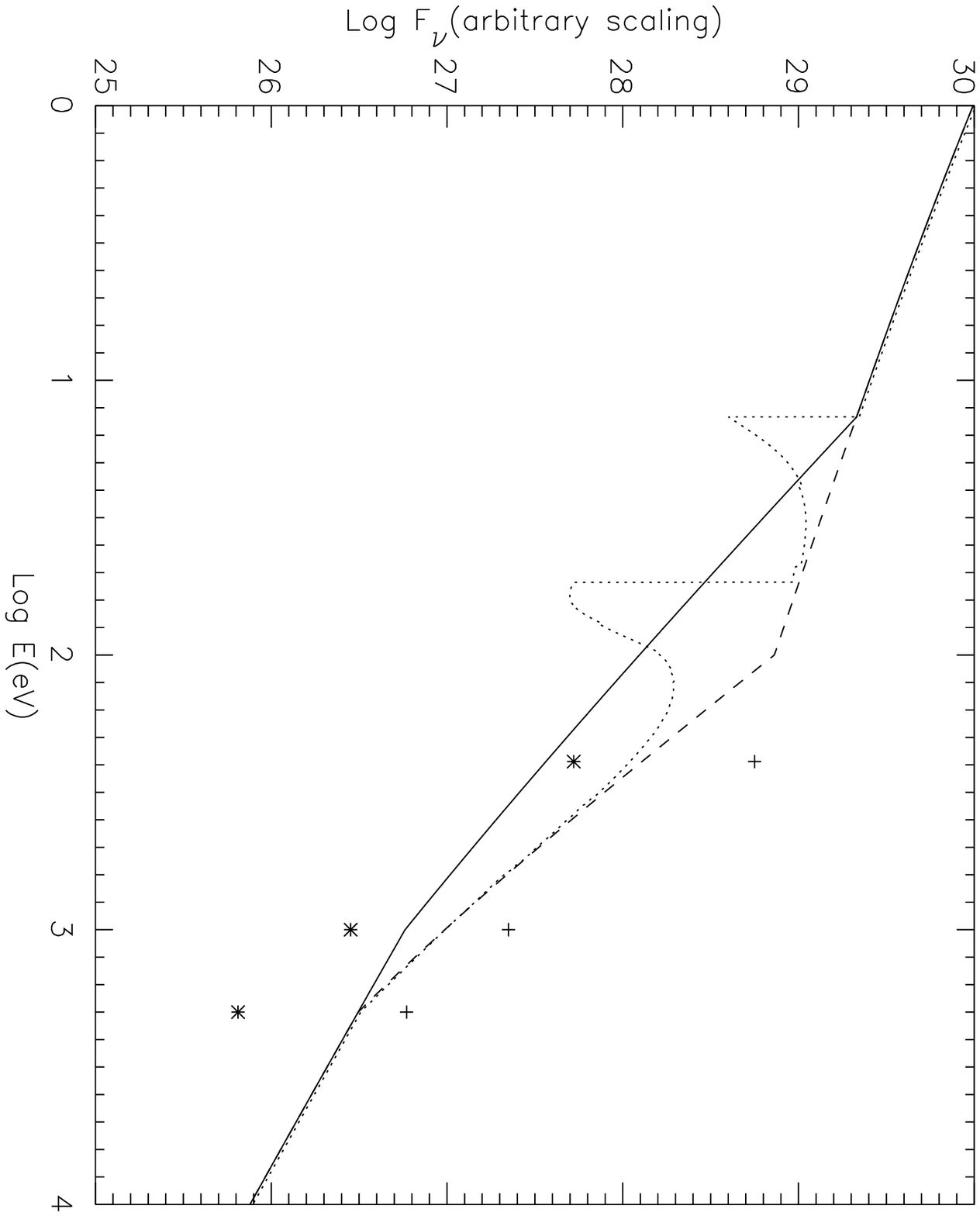]{A comparison of the model BLSy1 (dashed-line) and NLSy1
(solid line)
SEDs described in Section 2. The corrected luminosities at 0.25 keV, 1 keV, and 2 keV
are shown for Ton S180 (stars) and Ark 564 (crosses), from Turner et al. (2002) and Romano et al. 
(2004), respectively; note that the 0.25 keV point for Ark 564 was determined by extrapolation. The dotted line is the absorbed
continuum used to demonstrate the effect of an intervening absorber on the NLR gas in
NLSy1s.}

\figcaption[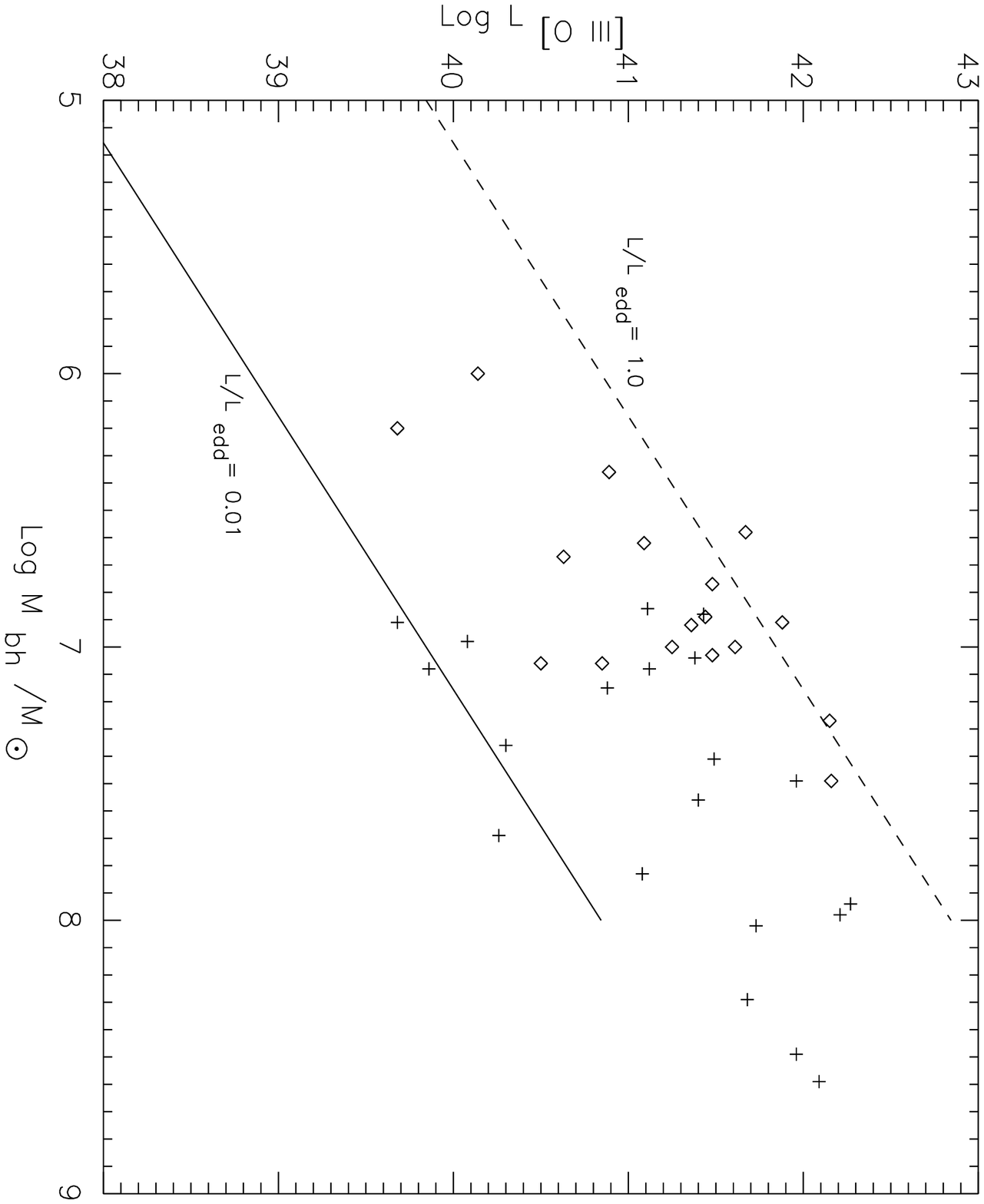]{The [O~III] $\lambda$5007 luminosity shown as 
a function of black hole mass for the Broad-Line and Narrow-Line
Seyfert galaxies in the sample (symbols as used in Figure 1). The luminosities are generally higher for the
NLSy1s, compared to BLSy1s of similar black hole mass, as expected if the former
have higher mass accretion rates. Also shown are the [O~III] luminosities as a function of black hole mass for 
Eddington ratios of 1.0 and 0.01 (dashed and solid lines, respectively).}

\clearpage
\vskip3.0in
\begin{figure}
\plotone{f1.ps}
\\Fig.~1.
\end{figure}

\clearpage
\vskip3.0in
\begin{figure}
\plotone{f2.ps}
\\Fig.~2.
\end{figure}

\clearpage
\vskip3.0in
\begin{figure}
\plotone{f3.ps}
\\Fig.~2.
\end{figure}

\clearpage

\begin{deluxetable}{lcccc}
\tablecolumns{5}
\scriptsize
\tablecaption{Broad-Line Seyfert 1 Galaxies}
\tablewidth{0pt}
\tablehead{
\colhead{Name} & \colhead{log(L$_{\rm [O~III]}$ ergs s$^{-1}$)}   
& \colhead{log(L$_{\rm 2-10 KeV}$ ergs s$^{-1}$)}
& \colhead{log(M$_{bh}$/M$_\odot$}) & \colhead{References$^{a}$}}
\startdata
3C 120 & 41.96 &  44.03 & 7.49 & 1,A\\
3C 390.3 & 42.09  & 43.93 & 8.59 & 2,A\\
Ark 120 & 41.68 & 43.80 & 8.29 & 3,A\\
Fairall 9 & 42.27 &  44.03 & 7.94 & 3,A \\
IC 4329A & 41.11 & 43.57 & 6.86 & 3,A\\
MCG-6-30-15 & 39.68 & 42.67 & 6.98 & 2,C\\
MCG 8-11-11 & 42.56 & 43.29 & & 3,\\
Mrk 79 & 41.73 & 43.35 & 8.02 & 3,A\\
Mrk 279 & 41.49 & 43.79 & 7.41 & 3,A\\
Mrk 290 & 41.51 & 43.24 & & 3\\
Mrk 509 & 42.21 & 44.07 & 7.98 & 2,A\\
Mrk 590 & 40.88 &   & 7.15 & 3,A\\
Mrk 704 & 41.09 & 42.96 &  & 2\\
Mrk 817 & 41.40 &  & 7.56 & 3\\
Mrk 841 & 41.96 & 43.52 & 8.49 & 2,B\\
Mrk 926 & 41.74 & 44.13 & & 2,\\
NGC 3227 & 40.26 & 42.02 & 7.69 & 3,A\\
NGC 3516 & 40.30 & 43.15 & 7.36 & 3,A\\
NGC 3783 & 41.38 & 42.95 & 7.04 & 2,A\\
NGC 4151 & 41.12 & 42.53 & 7.08 & 2,A\\
NGC 4593 & 40.08 & 42.67 & 6.91 & 2,B\\
NGC 5548 & 41.08 & 43.51 & 7.83 & 2,A\\ 
NGC 6814 & 39.86 & 40.76 & 7.08 & 3,B\\
NGC 7213 & 41.02 & 42.24 & & 2\\
NGC 7314 & 39.61 & 42.17 & & 2\\
NGC 7469 & 41.43 & 43.26 & 6.88 & 2,A\\
\enddata
\tablenotetext{a}{X-ray luminosities are averages from {\it ASCA} spectra in the
Tarturus database; except Mrk 79, which is from an {\it Ariel V} observation
(Elvis et al. 1978). The first value in the reference column is the source for the {O~III}  
luminosities, as follows: 1) Hua et al. 1988; 2) Nagao et al. 2000, and references therein;  
3) Bonatto \& Pastoriza 1997, and references therein. The 
second value is the source for the black hole mass: A) Wandel et al. 1999; B) Wandel 2002;
C) Czerny et al. 2002.}
\end{deluxetable}

\clearpage

\begin{deluxetable}{lcccc}
\tablecolumns{5}
\scriptsize
\tablecaption{Narrow-Line Seyfert 1 Galaxies}
\tablewidth{0pt}
\tablehead{
\colhead{Name} & \colhead{log(L$_{\rm [O~III]}$ ergs s$^{-1}$)}   
& \colhead{log(L$_{\rm 2-10 KeV}$ ergs s$^{-1}$)}
& \colhead{log(M$_{bh}$/M$_\odot$}) & \colhead{References$^{a}$}}
\startdata
Ark 564 & 41.44 & 43.33 & 6.89 & 2,D\\
IRAS 17020 & 41.48 & 43.82 & 6.77 & 3,F\\
I Zw 1 & 42.20 & 43.75 & & 2 \\
Mrk 42 & 40.14 & 42.21 & 6.00 & 2,E\\
Mrk 142 & 40.63 & 43.06 & 6.67 & 4,E\\
Mrk 110 & 41.88 & 43.89 & 6.91 & 2,A\\
Mrk 335 & 41.67 & 43.14 & 6.58 & 2,A\\
Mrk 478 & 42.15 & 43.54 & 7.27 & 3,F\\
Mrk 507 & 40.50 & 42.68 & 7.06 & 3,F\\
Mrk 705 & 41.36 & 43.27 & 6.92 & 2,E\\
Mrk 766 & 41.25 & 42.88 & 7.00 & 2,B\\
NGC 4051 & 39.68 & 41.55 & 6.20 & 3,A\\
PG 1011$-$040 & 41.48 & & 7.03 & 5,F\\
PG 1211$+$143 & 42.16 & 43.55 & 7.49 & 5,G\\
PG 1244$+$026 & 41.09 & 43.06 & 6.62 & 5,C\\
PG 1404$+$226 & 41.61 & 42.91 & 7.00 & 5,F\\
RE 1034$+$39 & 40.89 & 42.60 & 6.36 & 6,F\\
Ton S180 & 40.85 & 43.80 & 7.06 & 4,F\\
\enddata
\tablenotetext{a}{X-ray luminosities are from {\it ASCA} spectra in the
Tarturus database. References are as for Table 1, with the additions, as follows: 4) Grupe et al. 1998; 5) Miller et al. 1992;
6) Goncalves et al. 1999; D) Shemmer et al. 2001; E) Bian \& Zhao 2003.; F) Wang \& Lu 2001.
G) Kaspi et al. 2000.}
\end{deluxetable}

\end{document}